\documentclass[amssymb, floatfix, superscriptaddress, aps, prl, reprint]{revtex4-1}
\usepackage{graphicx}
\usepackage{dcolumn}
\usepackage{color}
\usepackage{amsmath}  %\text{} for text in math
\usepackage{bm}  %\bm{} for bold fo nt in math
\usepackage{hyperref}
\hypersetup{colorlinks=true, citecolor=blue, linkcolor=blue, urlcolor=blue}

\begin{document}
\title{Electron-Phonon Scattering in the Presence of Soft Modes \\ and Electron Mobility 
            in SrTiO$_{\text{3}}$ Perovskite from First Principles}
\author{Jin-Jian Zhou} 
\affiliation {Department of Applied Physics and Materials Science, California Institute of Technology, Pasadena, California 91125, USA}

\author{Olle Hellman} 
\affiliation {Department of Applied Physics and Materials Science, California Institute of Technology, Pasadena, California 91125, USA}
\affiliation{Department of Physics, Boston College, Chestnut Hill, Massachusetts 02467, USA}

\author{Marco Bernardi}
\email{bmarco@caltech.edu}
\affiliation {Department of Applied Physics and Materials Science, California Institute of Technology, Pasadena, California 91125, USA}
%\date{\today}

\begin{abstract}  \label{abstract}
Structural phase transitions and soft phonon modes pose a longstanding challenge to computing electron-phonon (e-ph) interactions in strongly anharmonic crystals. 
Here we develop a first-principles approach to compute e-ph scattering and charge transport in materials with anharmonic lattice dynamics. 
Our approach employs renormalized phonons to compute the temperature-dependent e-ph coupling for all phonon modes, including the soft modes associated with ferroelectricity and phase transitions. 
We show that the electron mobility in cubic SrTiO$_{3}$ is controlled by scattering with longitudinal optical phonons at room temperature and with ferroelectric soft phonons below 200~K. 
Our calculations can accurately predict the temperature dependence of the electron mobility in SrTiO$_{3}$ between 150$-$300~K, and reveal the microscopic origin of its roughly $T^{-3}$ trend. 
Our approach enables first-principles calculations of e-ph interactions and charge transport in broad classes of crystals with phase transitions and strongly anharmonic phonons.
\end{abstract}
\maketitle

\label{Intro:para1}  
Strontium titanate (SrTiO$_{\text{3}}$) is a prototypical perovskite oxide that has attracted interest due to its intriguing physical properties and technological applications~\cite{Hwang2012,Wang2016}. 
Similar to other perovskites, SrTiO$_{\text{3}}$ exhibits structural phase transitions (it is cubic above, and tetragonal below 105 K) with associated soft phonon modes that change their frequency with temperature~\cite{Scott1974,Cowley1969,Yamada1969}. 
This strongly anharmonic lattice dynamics is found broadly in materials of technological interest $-$ among others, metal-halide perovskites, oxides and chalcogenides. 
The complex interplay between electronic and lattice degrees of freedom makes it challenging to microscopically understand electron-phonon (e-ph) interactions and charge transport in these materials.\\
\indent
Despite extensive studies, the charge conduction mechanisms in SrTiO$_{\text{3}}$ are still debated~\cite{Frederikse1967,Wemple1969,Verma2014,Lin2017}. 
%is still incomplete, mainly due to the complex interplay the electronic and lattice degrees of freedom 
%The temperature dependence of charge transport and the underlying scattering mechanisms at intermediate temperature are yet to be understood~\cite{Frederikse1967,Wemple1969,Verma2014,Lin2017}.
%
The electron mobility in cubic SrTiO$_{\text{3}}$ exhibits a roughly $T^{-3}$ temperature dependence above 150~K~\cite{Lin2017,cain2013}, where carrier transport is typically limited by e-ph scattering.
However, it is still controversial whether the temperature dependence is due to scattering of electrons with longitudinal optical (LO) phonons, ferroelectric soft phonons~\cite{Frederikse1967,Wemple1969},  
or soft phonons associated with the cubic-to-tetragonal antiferrodistortive (AFD) phase transition~\cite{Verma2014,Zhou2016a}. %transverse optical (TO) 
While the arguments supporting each mechanism are based on phenomenological models, microscopic insight and quantitative analysis from first-principles calculations are still missing.\\
\indent %\label{Intro:para2}
Recently developed \emph{ab initio} calculations of e-ph coupling and phonon-limited carrier mobility~\cite{Mustafa2016,Zhou2016,Lee2018,Liu2017,Ma2018} are based on density functional perturbation theory (DFPT)~\cite{Baroni2001},
which cannot address strongly anharmonic lattice dynamics. 
% *** cite vanderbilt MAPbI paper?
Since DFPT predicts imaginary frequencies for the soft modes and lacks thermal effects, the typical workflow of \emph{ab initio} e-ph and charge transport calculations~\cite{Mustafa2016,Zhou2016,Lee2018}
cannot be applied to cubic SrTiO$_{\text{3}}$ and related materials with phase transitions and strong anharmonicity. %***footnote: the high-temperature phases*
Computing from first principles the scattering between electrons and soft phonons as a function of temperature remains an open challenge of broad relevance to materials physics.\\
%
% HERE WE SHOW
%
\indent %\label{Intro:para3}
In this Letter, we develop an \emph{ab initio} approach to compute the e-ph coupling as a function of temperature in strongly anharmonic crystals. 
We apply it to compute the phonon dispersions and the temperature dependence of the electron mobility in cubic SrTiO$_{\text{3}}$, obtaining results in excellent agreement with experiment. 
Our method allows us to quantify the contribution of different acoustic, optical and soft modes to e-ph scattering and transport. 
We find that both the AFD and the ferroelectric soft modes couple strongly with electronic states near the conduction band edge. We show that the $T^{-3}$ dependence of the mobility is due to an interplay between the LO and the ferroelectric soft phonons, which dominate e-ph scattering at temperatures above and below 200~K, respectively, while the AFD soft mode has a negligible contribution due to a lack of scattering phase space.
Our work provides a practical \emph{ab initio} approach to study e-ph coupling and charge transport in materials with anharmonic phonons, of which SrTiO$_{\text{3}}$ and related perovskite oxides are a paradigmatic case.\\ 
\indent  %\label{method}
The key ingredients for computing e-ph scattering and charge transport are the e-ph matrix elements $g_{mn\nu}\left(\bm{k},\bm{q}\right)$, which quantify the probability amplitude to scatter from an initial Bloch state $\left|n\bm{k}\right\rangle$ 
(with band $n$ and crystal momentum $\bm{k}$) to a final state $\left|m{\bm{k}+\bm{q}}\right\rangle$ by emitting or absorbing a phonon with wavevector $\bm{q}$, mode index $\nu$, energy $\hbar \omega_{\nu\bm{q}}$ and displacement eigenvector $\bm{e}_{\nu\bm{q}}$,
%
% EQUATION 1, g
%
\begin{equation}
g_{mn\nu}\left(\bm{k},\bm{q}\right)=\sqrt{\frac{\hbar}{2\omega_{\nu\bm{q}}}}\sum_{\kappa\alpha}\frac{\bm{e}_{\nu\bm{q}}^{\kappa\alpha}}{\sqrt{M_{\kappa}}}\left\langle m{\bm{k}+\bm{q}}\left|\partial_{\bm{q}\kappa\alpha}V\right|n\bm{k}\right\rangle,
\label{Eq:eph_mat}
\end{equation}
where $\partial_{\bm{q}\kappa\alpha}V\equiv \sum_{p} e^{i\bm{q}\bm{R}_{p}} \partial_{p\kappa\alpha}V$ and $\partial_{p\kappa\alpha}V$ 
is the variation of the Kohn-Sham potential for a unit displacement of atom $\kappa$ (with mass $M_{\kappa}$ and located in the unit cell at $\bm{R}_{p}$) in the Cartesian direction $\alpha$.\\
\indent
%
%***The temperature effect on the variation of self-consistent potential $\partial_{p\kappa\alpha}\bm{V}$ is neglected.\\
%
%\indent
To compute the e-ph coupling at finite temperature in anharmonic crystals, we use in Eq.~(\ref{Eq:eph_mat}) temperature-dependent renormalized phonon energies $\tilde{\omega}_{\nu\bm{q}}(T)$ and eigenvectors $\bm{\tilde{e}}_{\nu\bm{q}}(T)$ 
that include anharmonic effects and are obtained with the temperature-dependent effective potential (TDEP) method~\cite{Hellman2011,*Hellman2013a,*Hellman2013b}. 
TDEP extracts the effective interatomic force constants (IFCs) that best describe the anharmonic Born-Oppenheimer potential energy surface at a given temperature. For comparison, we also compute harmonic phonons and e-ph coupling using DFPT.
%  harmonic
For the TDEP calculations, we prepare a number of 4~$\times$~4~$\times$~4 (320 atom) supercells 
with random thermal displacements corresponding to a canonical ensemble at a given temperature $T$. 
%with random thermal displacements for a canonical ensemble at a given temperature $T$. 
To account for quantum fluctuations in the atomic positions, which are essential for the soft modes of SrTiO$_{3}$~\cite{Zhong1996}, the random displacements are generated using thermal amplitudes given by the Bose-Einstein distribution~\cite{Kim2018}.
%
%To account for the quantum fluctuations effect of atoms which is essential for the soft modes of SrTiO$_{3}$~\cite{Zhong1996}, the random displacements are generated using thermal amplitudes given by the Bose-Einstein distribution~\cite{Kim2018}. 
%
We then perform density functional theory (DFT) calculations (see below) on the supercells to collect atomic displacements and forces, and extract the effective force constants at each temperature by least-squares fitting~\cite{Hellman2011,*Hellman2013a,*Hellman2013b}. This process is repeated iteratively until convergence~\cite{Klein1972}.
% IS IT OKAY TO REMOVE THIS PART?***
%It should be noted that TDEP assumes a finite range of the IFCs (the IFCs are negligible outside the supercell), similar to the supercell approach for lattice dynamical calculations~\cite{Frank1995}. 
%For polar materials with non-vanishing Born effective charges, the long range dipole-dipole interactions violate this assumption~\cite{Parlinski1997,*Detraux1998,*Parlinski1998}. 
%
Note that in polar materials the IFCs contain a long-range contribution due to the dipole-dipole interactions~\cite{Gonze1994,Parlinski1997,*Detraux1998,*Parlinski1998}. 
We develop a new method to accurately include the long-range contribution in the IFCs; 
%treat the long-range dipole-dipole interactions due to long-wavelength polar phonons in supercells; 
the method is outlined in the Supplemental Material~\cite{ [{See \href{http://link.aps.org/supplemental/10.1103/PhysRevLett.121.226603}{Supplemental Material} for method to include the long-range interatomic force constants in polar materials in supercell calculations, computed bandstructure and effective masses in cubic SrTiO$_{3}$, and comparison of electron mobilities computed using different approaches, which includes Refs.~\cite{Frank1995,Giannozzi1991,Janotti2011}}] supp_mat} and will be detailed elsewhere.\\
\indent %Bernardi2014,Bernardi2015,
The e-ph matrix elements $g_{mn\nu}\left(\bm{k},\bm{q}\right)$ are computed, both using harmonic (DFPT) and anharmonic (TDEP) phonons, with our in-house developed {\sc perturbo} code~\cite{perturbo}, which is also employed to efficiently compute the e-ph scattering rates~\cite{Zhou2016} and the electron mobility using an iterative solution of the linearized Boltzmann transport equation (BTE)~\cite{Li2015}. 
%
%***The computation of the electron energies, wavefunctions and the perturbation potential $\partial_{p\kappa\alpha}V$ in Eq.~(\ref{Eq:eph_mat}) are performed using a workflow we previously employed~\cite{Zhou2016,Jhalani2017,Lee2018}. 
Briefly, we perform DFT calculations on SrTiO$_{3}$ within the Perdew-Burke-Ernzerhof generalized gradient approximation~\cite{Perdew2008} using the {\sc Quantum} {\sc Espresso} package~\cite{Giannozzi2009}. 
Fully relativistic norm-conserving pseudopotentials that include the spin-orbit coupling (SOC)~\cite{Hamann2013, Setten2018} are employed, together with 
the experimental lattice constant of 3.9{~\AA}~\cite{Loetzsch2010} and a plane-wave kinetic energy cutoff of 85 Ry. %The perturbation potential, $\partial_{p\kappa\alpha}V$ in Eq.~(\ref{Eq:eph_mat}), is obtained using DFPT~\cite{Baroni2001}. 
Wannier interpolation~\cite{Giustino2007} in combination with the polar correction~\cite{Sjakste2015,Verdi2015} is employed to evaluate the e-ph matrix elements on very fine Brillouin zone grids.  
We adopt coarse 8 $\!\times\!$ 8 $\!\times\!$ 8 $\bm{q}$-point grids for DFPT calculations, and Wannier functions for the Ti-$t_{2g}$ orbitals are constructed from 
Bloch states on a coarse 8~$\times$~8~$\times$~8 $\bm{k}$-point grid using the Wannier90 code~\cite{Mostofi2014}. 
Fine grids with up to 125$^{3}$ $\bm{k}$-points are used to converge the mobility.\\
\begin{figure}
\includegraphics[width=0.9\columnwidth]{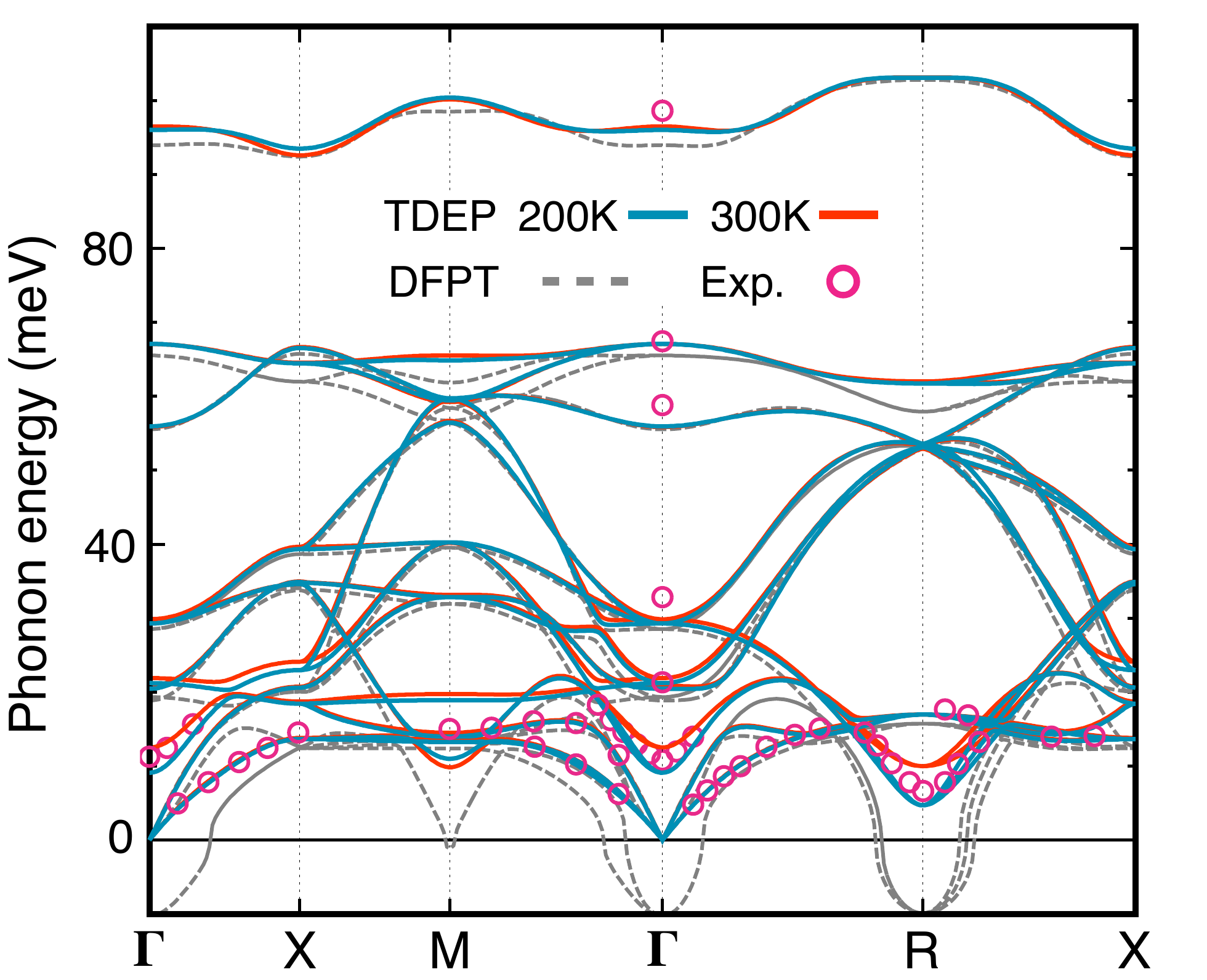}
\caption{Phonon dispersions of cubic SrTiO$_{3}$, computed with TDEP at 200~K (teal lines) and 300~K (red lines), and for comparison with DFPT (gray dashed line). 
Experimental results at room temperature (from Refs.~\cite{Stirling1972, Servoin1980}) are shown with open circles. 
%(d) The three lowest conduction bands of SrTiO$_{3}$, together with a color map of the electron mean free path.
}\label{Fig:fig1}
\end{figure}
\indent  %\label{results:phonon}
Figure~\ref{Fig:fig1} compares phonon dispersions computed with DFPT with those obtained using TDEP at 200 K and 300 K.
The DFPT result exhibits unstable soft phonon modes with negative energies both at the zone center ($\Gamma$ point) and corners ($R$ and $M$ points), consistent with previous work~\cite{Lebedev2009,Himmetoglu2014}. 
In the TDEP result, the soft phonons are stable, and their energy shifts continuously with temperature. 
Figure~\ref{Fig:fig1} also shows that the phonon dispersions obtained with TDEP at 300 K are in excellent agreement with experiment~\cite{Stirling1972,Servoin1980}.
The TDEP phonon dispersions at 200~K and 300~K match closely, except for the lowest-energy ferroelectric soft mode at $\Gamma$ and the soft AFD mode at $R$, 
for which energy renormalization due to anharmonic interactions is significant.
Although DFPT is inaccurate for the soft modes, it generates reasonable dispersions for high-energy phonons (above $\sim$30~meV) that are consistent with TDEP results.
The correction scheme introduced in this work~\cite{supp_mat} allows us to accurately account for the long-range contribution to the IFCs and to obtain accurate LO mode dispersions near $\Gamma$.
% due to polar phonons
%We note that an inadequate dipole-dipole correction would result in significant errors in the computed LO and TO dispersions near $\Gamma$. 
By contrast, recent work~\cite{Tadano2015,Feng2015} using the mixed-space approach~\cite{Wang2012} shows unusual oscillations along $\Gamma$$-$$R$ and $\Gamma$$-$$M$ in the highest LO mode dispersion, 
which are an artifact.\\
\indent
\begin{figure}
\includegraphics[width=0.9\columnwidth]{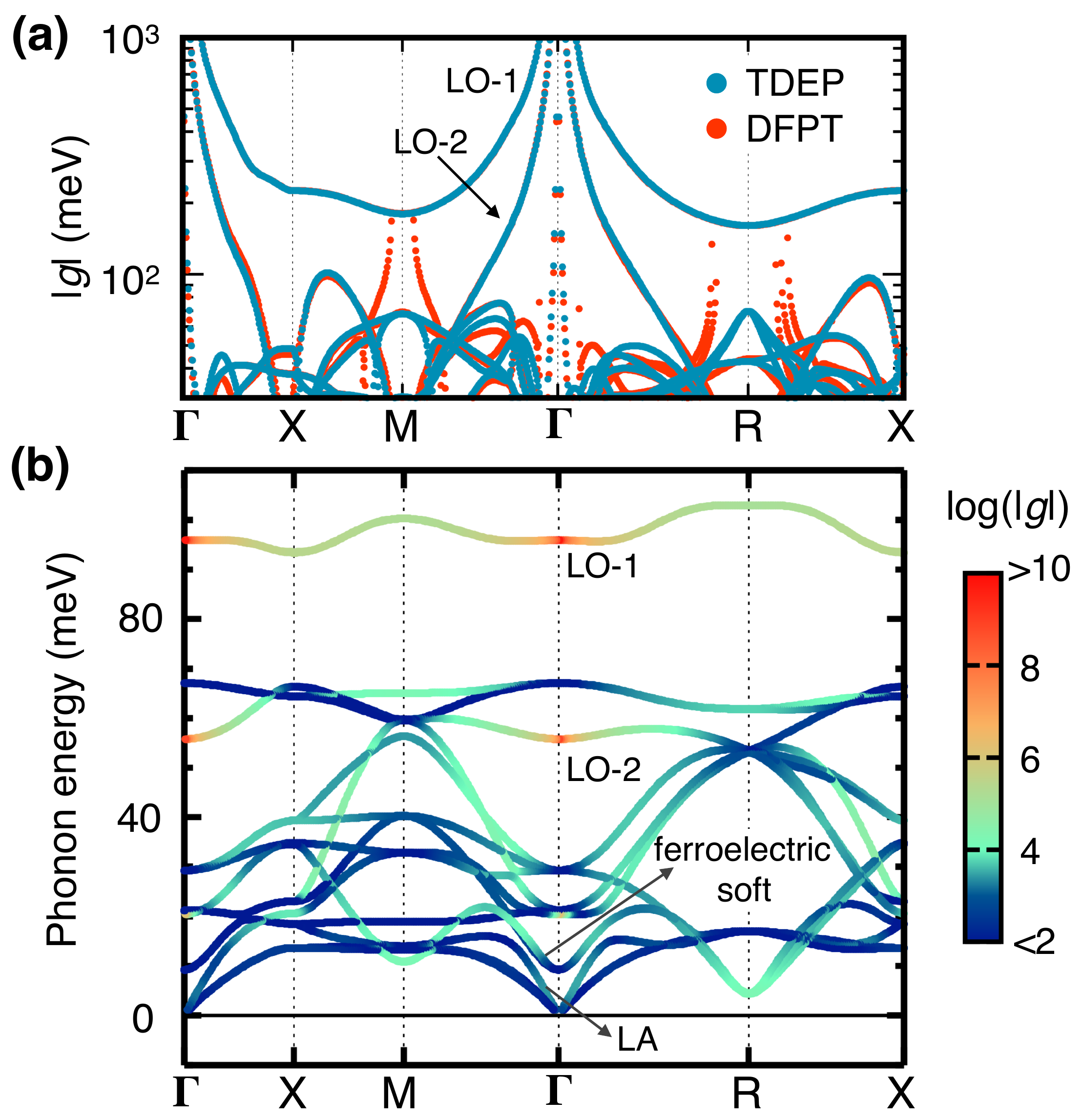}
\caption{(a) Absolute value of the e-ph matrix elements computed with anharmonic phonons from TDEP at 200~K (teal dots) and harmonic phonons from DFPT (red dots). 
%***We set the initial electron momentum $\bm{k}$ to be $\Gamma$, and compute $|g_{\nu}(\bm{q})|$ as the square root of the gauge-invariant trace of $|\bm{g}|^{2}$ over the three lowest conduction bands. 
(b) Phonon dispersions overlaid with a log-scale color map of $|g_{\nu}(\bm{q})|$.
}\label{Fig:fig2}
\end{figure}
To quantitatively study the coupling strength between electrons and different phonon modes, we analyze the absolute value of the e-ph matrix elements in Eq.~(\ref{Eq:eph_mat}), $|g_{mn\nu}(\bm{k},\bm{q})|$. 
We choose $\bm{k}\!=\!0$ (the $\Gamma$ point) as the initial electron momentum, and compute the square root of the gauge-invariant trace of $|g|^{2}$ over the three lowest conduction bands, for phonon wavevectors $\bm{q}$ along a high symmetry Brillouin zone path. Results are given for both anharmonic phonons computed at 200 K with TDEP and for harmonic phonons from DFPT for comparison.
The mode-resolved e-ph coupling strengths, $|g_{\nu}(\bm{q})|$, are shown in Fig.~\ref{Fig:fig2}(a) and mapped with a color scale on the phonon dispersions in Fig.~\ref{Fig:fig2}(b) for better visualization.
We find that the two highest-energy LO modes, labeled LO-1 and LO-2 in Fig.~\ref{Fig:fig2}(a,b), exhibit the strongest coupling with electrons; for these modes, 
$|g_{\nu}(\bm{q})|$ diverges as $1/q$ for $\bm{q}$ approaching $\Gamma$ due to the Fr\"ohlich interaction \cite{Froehlich1954}. 
Notably, both the ferroelectric soft mode near $\Gamma$ and the AFD soft mode at $R$ couple strongly with electrons.
While for the LO phonons the DFPT and TDEP results are in agreement, the coupling between electrons and soft modes exhibits an unphysical divergence in DFPT, whereas using TDEP anharmonic phonons gives a physical, finite value of $|g|$. 
The strong coupling between electrons and soft modes is essential to understanding electron dynamics in SrTiO$_3$.\\ %The methods introduced in this work allow us to correctly treat the soft modes and their coupling with electrons.\\
\begin{figure*}
\includegraphics[width=1.8\columnwidth]{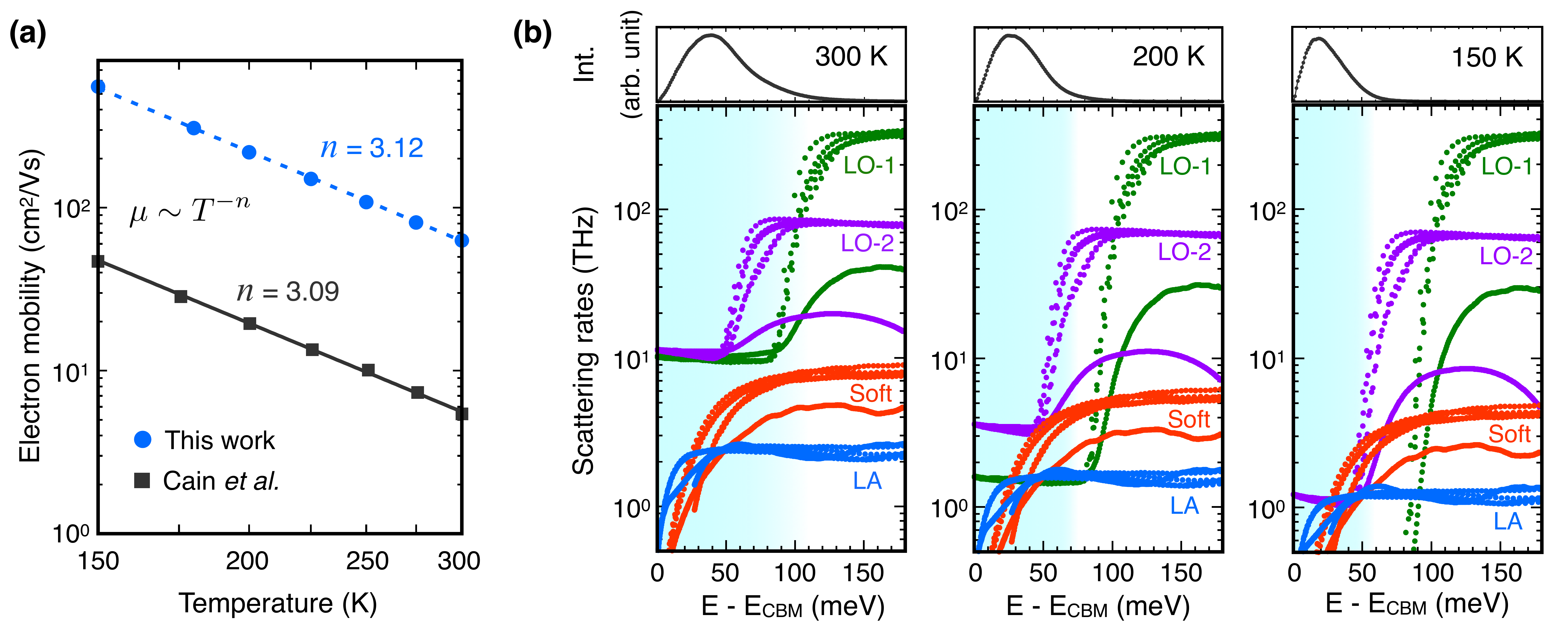}
\caption{ (a) Computed electron mobility as a function of temperature (blue circles), compared with experimental values (black squares) taken from Ref.~\cite{cain2013}. 
(b) Mode-resolved e-ph scattering rates, as a function of conduction band energy, at temperatures of 300~K, 200~K and 150~K (one per panel from left to right). 
The scattering rates are given for the two LO modes, the ferroelectric soft mode and the LA mode labeled in Fig.~\ref{Fig:fig2}(b). 
For better visualization, only the scattering rates of electronic states with $\bm{k}$ along $\Gamma-X$ and $\Gamma-M$ are shown, which correspond to the lower and upper bounds of the $\bm{k}$-dependent scattering rates. The integrand in Eq.~(\ref{eq:cond}), which shows how much electronic states at a given energy contribute to transport, is also plotted at each temperature. The zero of the energy axis is the CBM.}
\label{Fig:fig3}
\end{figure*}
\indent  %\label{results:t-dep}
%
% MOBILITY
%
The temperature-dependent e-ph matrix elements are employed to compute the mobility using both the relaxation time (RT) approximation~\cite{Zhou2016} and an iterative solution of the BTE~\cite{Li2015} that goes beyond the RT approximation. 
Briefly,  we compute the e-ph scattering rates (and their inverse, the RTs $\tau_{n\bm{k}}$) from the imaginary part of the lowest-order e-ph self-energy~\cite{Bernardi2016}. The mobility is computed as 
\begin{equation}
\mu_{\alpha\beta}(T) = \frac{2e}{n_{c}V_{\text{uc}}} \int{{\!dE} \left(\!-\frac{\partial{f}}{\partial{E}} \right)\!\sum_{n\bm{k}}{ \bm{F}_{n\bm{k}}^{\alpha}(T) \bm{v}_{n\bm{k}}^{\beta}\delta(E-\varepsilon_{n\bm{k}}) }},
\label{eq:cond}
\end{equation}
where $\varepsilon_{n\bm{k}}$ and $\bm{v}_{n\bm{k}}$ are the electron energy and velocity, respectively, $\alpha$ and $\beta$ are Cartesian directions, $f$ is the Fermi-Dirac distribution, $V_{\text{uc}}$ is the unit cell volume and $n_{c}$ is the electron concentration.  
$\bm{F}_{n\bm{k}}$ is computed as $\tau_{n\bm{k}}\bm{v}_{n\bm{k}}$ in the RT approximation, or obtained by solving the BTE iteratively.
We perform mobility calculations at different $n_{c}$ ranging between $10^{17}-10^{19}~\text{cm}^{-3}$, with the chemical potential obtained numerically for each carrier concentration.\\
%We perform mobility calculations at different $n_{c}$ ranging between $10^{17}~\text{cm}^{-3}$ and $10^{19}~\text{cm}^{-3}$ in the rigid band shift approximation and the chemical potential is determined accordingly.\\
\indent
Figure~\ref{Fig:fig3}(a) shows our calculated electron mobility as a function of temperature (obtained using the iterative BTE solution) and compares it with experimental measurements from Ref.~\cite{cain2013}. 
Since electron-defect scattering is neglected in our mobility calculations, we compare our results to experimental data above 150~K, where the mobility is nearly independent of carrier concentration; note that an interplay between disorder and electron-phonon scattering, which is relevant in some experiments, cannot be ruled out. 
%
%and at the lowest carrier concentration ($n_{c}$=$8 \times 10^{17}~\text{cm}^{-3}$), where the contribution of electron-defect scattering to the electron mobility is negligible.
%
The temperature dependence of our computed mobility is in excellent agreement with experiment. By fitting the data at 150$-$300~K with a $T^{-n}$ power law, 
we get $n\!\approx\!3.09$ for the experimental data and $n\!\approx\!3.12$ for our computed mobility, namely an error in the exponent within 1\%.
The RT approximation and the iterative BTE solution exhibit a similar temperature dependence of the mobility [see Fig.~S2(a)]~\cite{supp_mat}. In particular, the iterative BTE mobility at room temperature is roughly 15\% higher than the mobility in the RT approximation, and its temperature dependence of $T^{-3.1}$ agrees better with experiment than the RT approximation result ($T^{-3.3}$). Our results clearly show that the $T^{-3}$ dependence of the mobility can be explained through the e-ph scattering alone. \\
%, ruling out alternative explanations~\cite{Lin2017}. 
%
% INTERPLAY OF DIFFERENT PHONON MODES ON MOBILITY
%
\indent
There is a subtle interplay between the e-ph scattering mechanisms regulating the temperature dependence of the electron mobility. 
We analyze the e-ph scattering rates for each phonon mode and their contribution to transport, highlighting the role of the soft modes. 
% near $\Gamma$ which
We focus on the four modes labeled in Fig.~\ref{Fig:fig2}(b) $-$ the two LO modes, the longitudinal acoustic (LA) mode and the ferroelectric soft mode near $\Gamma$ $-$ that are most relevant for transport at 150$-$300~K.
Note that, although the AFD soft mode at $R$ has an even stronger e-ph coupling than the ferroelectric soft mode [see Fig.~\ref{Fig:fig2}(b)], it does not scatter electrons appreciably due to a lack of scattering phase space near the CBM (see Fig.~S1)~\cite{supp_mat}, and thus gives a negligible contribution to transport.
Figure~\ref{Fig:fig3}(b) shows the mode-resolved scattering rates at three representative temperatures.
%
% TDF
% 
Also shown in Fig.~\ref{Fig:fig3}(b) is the integrand in Eq.~(\ref{eq:cond}) at each temperature, which quantifies how much electronic states at a given energy contribute to transport~\cite{Zhou2016}. 
Between 150$-$300 K, the integrand is non-zero only within $\sim$100 meV of the conduction band minimum (CBM), so that the e-ph scattering rates in that energy range can accurately quantify which phonon modes limit the mobility.\\
\indent
At 300~K, the two LO modes dominate e-ph scattering, exhibiting scattering rates an order of magnitude higher than any other mode. The LO-mode scattering rate as a function of energy has a two-plateau structure ~\cite{Zhou2016,Lee2018}; the low-energy plateau, which mainly contributes to transport, corresponds to LO phonon absorption, a thermally activated process with a rate proportional to the LO phonon occupation, $N_{\rm{LO}}\!\approx\! e^{-\hbar \omega_{\rm{LO}}/kT}$. 
As the temperature is reduced, the contribution from LO mode scattering thus drops exponentially and at lower temperatures transport is dominated by scattering with low-energy phonons, including acoustic and soft modes.\\
\indent
At 200~K, scattering from the ferroelectric soft mode is significant. Its rate is larger than the scattering rate with the LO-1 mode, and it is second only to scattering with the LO-2 mode.  
At 150~K, ferroelectric soft mode scattering is dominant in the energy range of interest for transport, with a smaller contribution from scattering by LA phonons. 
Our results show that, while LO phonon scattering limits the mobility at room temperature, the ferroelectric soft phonons play a crucial role at lower temperatures, dominating over other scattering mechanisms near 150~K.
The $T^{-3}$ mobility dependence is due to the combined effects of the LO mode and ferroelectric soft mode scattering.
%and it is not due to a single phonon mode as hypothesized in previous work~\cite{Frederikse1967,Wemple1969,Himmetoglu2014}. 
%
%Although the coupling between electrons and the AFD soft mode at $R$ is also strong [see  Fig.~\ref{Fig:fig2}(b)], the AFD soft mode does not scatter electrons appreciably due to a lack of scattering phase space near the CBM. 
%
For comparison, the electron mobility computed using the DFPT phonons, in which the contribution from the ferroelectric soft mode is absent, shows a much stronger temperature dependence, roughly $T^{-4}$ as shown in Fig.~S2(b)~\cite{supp_mat}.
Our accurate treatment of soft phonons and their temperature-dependent e-ph scattering is crucial to gain these new microscopic insights.\\
%Although our analysis focuses on cubic SrTiO$_3$, 
%We expect that in the tetragonal phase (below 105 K) the AFD soft phonon plays an important role in transport since it is folded to $\Gamma$, thus acquiring a large e-ph scattering phase space. 
%
%only small $\bm{q}$ phonon scattering is possible (see Fig.~\ref{Fig:fig1}(d)).
%However, if temperate go below $T_{c}$, the AFD soft phonon modes will folded into $\Gamma$-point due to the cubic-to-tetragonal structural phase transition. 
%Hence, we expect AFD soft phonon to play an important role at temperature blow $T_{c}$. 
%down folding, the AFD soft phonon start to enter,  this might be account the results in electron-soft phonon scattering. \\
%
\indent
%We assume band-like transport in SrTiO$_{3}$ and compute the mobility using an accurate iterative solution of BTE.% beyond RT approximation. 
As seen in Fig.~\ref{Fig:fig3}(a), the computed mobility is almost an order of magnitude higher than experiments. We maintain that our results are accurate within the band-like picture of transport and the lowest-order of perturbation theory in the e-ph interaction. 
Note that our calculations include SOC effects and a bandstructure with accurate electron effective masses~\cite{supp_mat}, use phonons in excellent agreement with experiments (including the soft modes), treat scattering from all phonon modes on the same footing, are carefully converged using ultra-fine grids, and employ an accurate iterative solution of BTE to obtain the mobility. 
We found previous work that reported \textit{ab initio} calculations of the room temperature electron mobility of SrTiO$_3$ in good 
agreement with experiment~\cite{Himmetoglu2014}. In that work, the mobility was computed within the RT approximation, without including SOC effects or soft modes, and by including only LO phonon scattering with an approximate treatment. 
%that is less accurate than the \textit{ab initio} Frohlich formula. 
Our results show that accurate calculations within \emph{ab initio} band theory and the lowest-order e-ph interaction significantly overestimate the electron mobility in SrTiO$_{3}$.\\
%, thus invalidating these previous results.
%However, our transport calculations, which takes into account the spin-orbital coupling effect on the electronic structures, takes all the phonon scattering on the same footing including the soft phonon scattering, and , show a much greater electron mobility. 
%Here we do a more throughout calculations, where we include all the phonon modes including the soft phonon modes, spin orbital coupling, extremely fine grid, and iterative solution. 
%
\indent %strong e-ph interactions and 
We argue that the discrepancy between the measured and the \textit{ab initio} band-like mobility is due to polaron effects~\cite{Devreese2009}, 
which are known to occur in oxide crystals like SrTiO$_3$ with mobilities of less than $\sim$10~cm$^2$/V$\,$s at room temperature~\cite{Emin2012}. 
The presence of large polarons in SrTiO$_{3}$ is well-known experimentally~\cite{Mechelen2008,Chang2010}. 
A theory that includes strong e-ph interactions beyond the lowest order is needed to more accurately compute the mobility in the polaron transport regime. 
One expects that including higher-order e-ph interactions would suppress the e-ph RTs and lower the computed mobility towards the experimental value. 
While developing an \textit{ab initio} theory of polaron transport will be the subject of future work, it is clear that the temperature dependence of the mobility agrees well with experiment within our lowest-order approach.\\
% it is clear that 
%
%*** In addition, our computed e-ph coupling strength $|g(\bm{q})|$ for the highest LO mode in SrTiO$_{3}$ (see Fig.~\ref{Fig:fig2}) are comparable to those reported for anatase TiO$_{2}$ (see Fig.~\ref{Fig:fig1} of Ref.~\cite{Verdi2015}), where the presence of large polarons has been confirmed~\cite{Moser2013,Verdi2017}. This result indicates that e-ph coupling is strong enough in SrTiO$_{3}$ to induce formation of a large polaron. 
%
%Future work will investigate large polaron transport from first principles.\\%by including higher orders in the e-ph interaction. \\
%
%in  has also been discussed based on  and theoretical model 
%We attribute the discrepancy to the polaronic effect that we don't includes.
%our computed g is compared to that of TiO2, which have found to large polaron, 
%experimental data show controversial results. 
%further work should explore the polaron transport. 
%
\indent  %\label{conclusion}
In summary, we developed a first-principles approach to compute e-ph interactions and charge transport in materials with phase transitions and anharmonic lattice dynamics.
Accurately treating the soft modes reveals the origin of the $T^{-3}$ temperature dependence of the electron mobility in cubic SrTiO$_{3}$, %the dominant scattering mechanisms and 
which we show to be due to the combined e-ph scattering from LO and soft ferroelectric modes. 
%
%Our approach enables first-principles calculations of e-ph interactions in materials with structural phase transitions and anharmonic lattice dynamics. 
Our work paves the way to studying charge carrier dynamics in broad classes of materials with anharmonic phonons, including perovskite oxides, metal-halide perovskites and
chalcogenides. 
%SnSe and related compounds.  % SnSe
\\
%We apply our approach to cubic SrTiO$_{3}$ where we compute the temperature-dependent phonon dispersion and electron mobility. Our computed mobility presents a nearly cubic-power-law dependence on temperature that is in excellent agreement with experiments.
%We further analyze the electron scattering by different phonon modes, and demonstrate the important role of ferroelectric soft phonon scattering in carrier transport of SrTiO$_{3}$ at intermediate temperature. 
%Contrary to previous calculations, our computed free carrier mobility at room temperature is one order of magnitude larger than experimental values, which indicates that polaronic effect dominates the transport properties in SrTiO$_{3}$ at room temperature. Our work provide a powerful tool to investigate microscopic process in many technological relevant materials, such as oxides, thermoelectric materials.
%
\newline
\indent \label{Acknowledge}
This work was supported by the Joint Center for Artificial Photosynthesis, a DOE Energy Innovation Hub, supported through the Office of Science of the U.S. Department of Energy under Award No.~DE-SC0004993. 
M.B. acknowledges support by the National Science Foundation under Grant No.~ACI-1642443, which provided for code development, and Grant No.~CAREER-1750613, which provided for theory and method development. 
This work was partially supported by the Air Force Office of Scientific Research through Young Investigator Program Grant FA9550-18-1-0280.
O.H. acknowledges support from the EFRI-2DARE program of the National Science Foundation, Award No. 1433467. 
This research used resources of the National Energy Research Scientific Computing Center, a DOE Office of Science User
Facility supported by the Office of Science of the U.S. Department of Energy under Contract No. DE-AC02-05CH11231.
\bibliographystyle{apsrev4-1}
\bibliography{sto_ref}
\end{document}